\documentclass{aa}
\usepackage{graphicx}
\usepackage{txfonts}
\usepackage[colorlinks=true, linkcolor=blue, citecolor=blue, filecolor=blue, urlcolor=blue]{hyperref}
\usepackage{CJKutf8}
\begin{document} 
\title{Sequential giant planet formation initiated by disc substructure}

\author{Tommy Chi Ho Lau
\begin{CJK*}{UTF8}{bsmi}(劉智昊)\end{CJK*}
\inst{1}
\and
Til Birnstiel\inst{1,2}
\and
Joanna Dr{\c a}{\.z}kowska\inst{3}
\and
Sebastian Markus Stammler\inst{1}
}

\institute{University Observatory, Faculty of Physics, Ludwig-Maximilians-Universität München, Scheinerstr. 1, 81679 Munich, Germany
\label{inst1}
\and
Exzellenzcluster ORIGINS,Boltzmannstr. 2, Garching, 85748, Germany
\label{inst2}
\and
Max Planck Institute for Solar System Research, Justus-von-Liebig-Weg 3, Göttingen, 37077, Germany
\label{inst3}
}

\date{Received 22 April 2024; accepted 17 June 2024}
 
\abstract
{Planet formation models are necessary to understand the origins of diverse planetary systems. Circumstellar disc substructures have been proposed as preferred locations of planet formation, but a complete formation scenario has not been covered by a single model so far.}
{We aim to study the formation of giant planets facilitated by disc substructure and starting with sub-micron-sized dust.}
{We connect dust coagulation and drift, planetesimal formation, $N$-body gravity, pebble accretion, planet migration, planetary gas accretion, and gap opening in one consistent modelling framework.}
{We find rapid formation of multiple gas giants from the initial disc substructure. The migration trap near the substructure allows for the formation of cold gas giants. A new pressure maximum is created at the outer edge of the planetary gap, which triggers the next generation of planet formation resulting in a compact chain of giant planets. A high planet formation efficiency is achieved, as the first gas giants are effective at preventing dust from drifting further inwards, which preserves material for planet formation.}
{Sequential planet formation is a promising framework to explain the formation of chains of gas and ice giants.}

\maketitle
\section{Introduction} \label{sec:intro}
Planet formation is a multi-step process spanning over 40 orders of magnitude in mass. In recent years, there has been significant progress in the understanding of this process driven by the numerous discoveries of exoplanets and observations of protoplanetary discs (see \cite{Drazkowska2023} for a recent review). The formation of the first gravitationally bound building blocks of planets, the planetesimals, which used to be a major bottleneck of the planet formation theory, was addressed by the streaming instability \citep{Johansen2007}. The properties of planetesimals formed in the streaming instability broadly match comets and the Kuiper belt objects \citep{Blum2017, Nesvorny2019}. The growth timescale of giant planet cores, which was prohibitively long in the classical planetesimal-driven paradigm \citep{Lissauer1987,Kokubo2000,Kokubo2002}, has been addressed by introducing pebble accretion \citep{Ormel2010, Lambrechts2014}. Despite these new developments, a consistent model covering all the stages of planet formation does not exist yet. Formation of the cores of giant planets remains a challenge, particularly at large orbital distances \citep{Voelkel2020, Coleman2021, Eriksson2023}.

 
Current planet formation models in general fail to meet both the physical and cosmochemical constraints required to explain the Solar System \citep[e.g.][]{Matsumura2017,liu2019,Bitsch2019,Lau2024}. The main challenge remains to be the `migration problem', where rapid migration occurs for planetary cores of 1 to 10 $M_\oplus$, resulting in the formation of super-Earths and hot Jupiters.
While the above models generally assume a smooth planetary disc, multiple works \citep[e.g.][]{coleman2016,morbidelli2020,guilera2020,Chambers2021,andama2022,Lau2022} have modelled the formation and evolution of planetary cores retained at the migration trap near a pressure bump in the disc.
Nonetheless, the origin of such pressure bumps remains uncertain, the proposed non-planetary causes include late-stage infall of material \citep{gupta2023}, sublimation \citep{saito2011}, instabilities \citep{takahashi2014,flock2015,dullemond2018a}, and the edge of the dead zone, where the magneto-rotational instability (MRI) is suppressed \citep{pinilla2016}.

More recently, the high-resolution interferometry observations by the Atacama Large Millimeter/submillimeter Array (ALMA) have shown that substructures are typical in protoplanetary discs. Multiple surveys \citep[e.g.][]{andrews2018,long2018,cieza2021}, have shown that most of the substructures are presented in the form of axisymmetric rings. While these observations are limited to large and bright discs, disc population synthesis and theoretical models \citep{toci2021,zormpas2022} have demonstrated that disc substructures may be common in unresolved discs as well.

\cite{chatterjee2013} presented an analytic model demonstrating the `inside-out' planet formation scenario, where planet formation starts at the outer edge of the MRI active zone around the star. Although this work focuses on explaining the tightly packed chains of planets commonly seen in exoplanetary systems, the model suggests the possibility of planet formation being triggered by the planet formed in the previous generation. A similar idea was also proposed for the formation of Saturn after the completion of Jupiter \citep{kobayashi2012}, in which the core of Saturn grows rapidly without significant inward drift in the pressure bump induced by the planetary gap of Jupiter, although they did not consider planetesimal formation from dust and pebble accretion. 

Motivated by the current models and observations, a substructure in a protoplanetary disc has recently emerged as an ideal location for planet formation \citep{Chambers2021,Lau2022,Jiang2023}. \cite{Lau2022} modelled the formation and evolution of planetesimals in an initial axisymmetric disc substructure by coupling the dust and gas evolution code DustPy \citep{Stammler2022} with the parallelised symplectic $N$-body integrator SyMBA parallelised (SyMBAp; \citealp{Lau2023}). As the disc evolves and satisfies the condition for the streaming instability and the subsequent gravitational collapse, a fraction of the dust is converted into planetesimals as $N$-body particles. On top of the full $N$-body gravitational interactions, additional subroutines are gas drag, planet-disc interactions, and pebble accretion. \cite{Lau2022} showed the rapid formation of planetary cores thanks to the concentrated dust, which are also retained due to the migration trap near the pressure bump. However, this work did not attempt to form giant planets, as the authors focused on the formation of planetary cores, where gas accretion and planetary gap opening are missing in the model.

In this work, we further developed the model in \cite{Lau2022} for the formation of giant planets initiated by a disc substructure and found a scenario of sequential planet formation. In the following, Sec. \ref{sec:method} summarises the methods adopted in \cite{Lau2022} and the new components implemented in this work. The results are presented in Sect. \ref{sec:results}, which are followed by the discussions in Sect. \ref{sec:dis}. The findings of this work are summarised in Sect. \ref{sec:conc}.

\section{Method} \label{sec:method}
We employed the dust and gas evolution code DustPy v1.0.3 \citep{Stammler2022} and the symplectic $N$-body integrator SyMBAp v1.6 \citep{Lau2023}, a parallelised version of the Symplectic Massive Body Algorithm (SyMBA; \citealp{duncan1998}). The coupling of the two codes to construct a consistent planet formation model was first presented in \cite{Lau2022}, which only modelled the formation of planetary cores. In this work, we added gas accretion and planetary gap opening to model the subsequent evolution of the embryos formed at a pressure bump. The following summarises the employed method in \cite{Lau2022} and describes the new components in detail.

\subsection{Disc model}
DustPy simulates the viscous evolution of the gas, coagulation, fragmentation, advection, and diffusion of the dust in a protoplanetary disc. The different parts of the disc model are described in this section.

\subsubsection{Gas component}
We considered a protoplanetary disc around a solar-type star, which is axisymmetric and in vertical hydrostatic equilibrium. The initial gas surface density $\Sigma_{\mathrm{g,init}}$ is given by
\begin{equation}
	\Sigma_{\mathrm{g,init}} = \frac{M_\mathrm{disc}}{2\pi r_\mathrm{c}^2}\left( \frac{r}{r_\mathrm{c}}\right) ^{-1}\exp\left(-\frac{r}{r_\mathrm{c}}\right),
\end{equation}
with the distance from the star $r$, the initial mass of the disc $M_\mathrm{disc}$, and the characteristic radius $r_\mathrm{c}$. We set $M_\mathrm{disc}=0.0263M_\odot$ and $r_\mathrm{c}=50\ \mathrm{au}$, which imply  $\Sigma_{\mathrm{g,init}} \ (r=5\ \mathrm{au})\approx 134.6\ \mathrm{g}\ \mathrm{cm}^{-2}$.

The gas disc viscously evolves in time $t$ according to the advection-diffusion equation
\begin{equation}
	\frac{\partial \Sigma_\mathrm{g}}{\partial t}=\frac{3}{r}\frac{\partial}{\partial r}\left[ r^{1/2} \frac{\partial}{\partial r}(\nu \Sigma_\mathrm{g} r^{1/2}) \right]
\end{equation}
\citep{Lust1952,Lynden-Bell1974}, while the backreaction from the dust is neglected. The \cite{Shakura1973} $\alpha$-parametrisation was adopted for the kinematic viscosity $\nu$ such that
\begin{equation}
	\nu = \alpha c_\mathrm{s} H_\mathrm{g},
\end{equation}
with the speed of sound $c_\mathrm{s}$ and the disc scale height $H_\mathrm{g}$. The viscosity parameter $\alpha=\{3,5\}\times10^{-4}$ was set in this work. The disc scale height is defined by $H_\mathrm{g}\equiv c_\mathrm{s}/\Omega_\mathrm{K}$, where the local Keplerian orbital frequency $\Omega_\mathrm{K}=\sqrt{GM_\odot/r^{3}}$ with the gravitational constant $G$. The isothermal sound speed was used and given by $c_\mathrm{s} = \sqrt{k_\mathrm{B}T/\mu}$ with the Boltzmann constant $k_\mathrm{B}$, the midplane temperature $T$ and the mean molecular weight of the gas $\mu=2.3m_\mathrm{p}$. The disc was assumed to be passively irradiated by the solar luminosity at a constant angle of 0.05, which gives a midplane temperature profile
\begin{equation}
	T\approx221 \left( \frac{r}{\mathrm{au}}\right) ^{-1/2} \mathrm{K}.
\end{equation}
We note that the normalisation is about 0.84 of that in \cite{Lau2022}, which is the result of a correction made to DustPy since v1.0.2.
This setup yields the dimensionless gas disc scale height
\begin{equation}
	\hat{h}_\mathrm{g}\equiv\frac{H_\mathrm{g}}{r}\approx0.0299\left( \frac{r}{\mathrm{au}}\right)^{1/4}.
\end{equation}
The midplane pressure gradient parameter $\eta$ is then given by
\begin{equation}
	\eta=-\frac{\hat{h}_\mathrm{g}^2}{2}\frac{\partial\ln P}{\partial\ln r}, \label{eq:eta}
\end{equation}
with the midplane gas pressure $P$, which describes the degree of `sub-Keplerity' of the gas. A logarithmic radial grid was adopted with with 133 cells from 3 to 53 au and with an additional 42 cells from 53 to 1000 au.

\subsubsection{Dust component}
The initial dust surface density $\Sigma_{\mathrm{d,init}}$ is given by
\begin{equation}
	\Sigma_{\mathrm{d,init}} = Z\Sigma_{\mathrm{g,init}}
\end{equation}
with the global dust-to-gas ratio $Z$ set at the solar metallicity of 0.01. We followed \cite{Mathis1977}, that is, the MRN size distribution of the interstellar medium, for the initial size distribution of the dust grains. The maximum initial size was set at $1\ \mu\mathrm{m}$ with the internal density of $1.67\ \mathrm{g}\ \mathrm{cm}^{-3}$ assumed. A total of 141 dust mass bins logarithmically spaced from $10^{-12}$ to $10^8$ g were used. Each dust species was evolved in time through transport with the advection-diffusion equation \citep{Clarke1988} coupled to growth and fragmentation with the Smoluchowski equation. The fragmentation velocity was assumed to be $5\ \mathrm{m}\mathrm{s}^{-1}$. At collision velocities above which, the dust aggregates are assumed to fragment. The Stokes number St$_i$ of the dust in each dust species $i$ was calculated by considering the Epstein and the Stokes I regimes. The dust scale height of each dust species $H_{\mathrm{d},i}$ was calculated according to \cite{Dubrulle1995},
\begin{equation}
	H_{\mathrm{d},i}=H_\mathrm{g}\sqrt{\frac{\alpha}{\alpha+\mathrm{St}_i}}, \label{eq:h_d}
\end{equation}
assuming $\mathrm{St}_i<1$.
Further details of the algorithms for the disc model are described in \cite{Stammler2022}.

\subsubsection{Initial disc gap} \label{sec:gap}
An initial axisymmetric gap was introduced to the disc following the model by \cite{Dullemond2018}, which is motivated by the commonly observed substructures in protoplanetary discs. To modify the gas profile, we applied a modified $\alpha$-parameter with radial dependence $\alpha'(r)= \alpha/F(r)$,
where the function
\begin{equation}
	F(r)=\exp\left[ -A \exp\left(-\frac{(r-r_0)^2}{2w^2} \right) \right]\label{eq:F_gap}
\end{equation}
with the gap amplitude $A=1$, the location $r_0=5.5$ au and the gap width $w=0.5$ au. The initial gap is removed when the first planet has the gap opening factor $K$ \citep{Kanagawa2015} of 250. The $K$ factor is further described in Sect. \ref{sec:method:K} on planetary gap opening. This value translates to a perturbation towards the gas disc of about $1/10$ of the unperturbed gas surface density.

Since we do not study the physical cause of the initial disc gap, the modified $\alpha'(r)$-parameter only serves the purpose to attain a target gas profile and does not change the actual turbulence in the disc.
Therefore, the modified $\alpha$-parameter $\alpha'(r)$ is exclusively experienced by the gas, while dust diffusion, dust scale height and turbulent collision speeds are set by $\alpha$. Nonetheless, the dust evolves according to the resulting gas profile. We note that this treatment is not consistent with the substructure formation scenarios where the actual turbulence is changed, for example, the edge of the dead-zone, but consistent with the cases where the turbulence is unchanged, for example, infall of material.

\subsubsection{Planetesimal formation}
\begin{figure}
	\centering
	\includegraphics{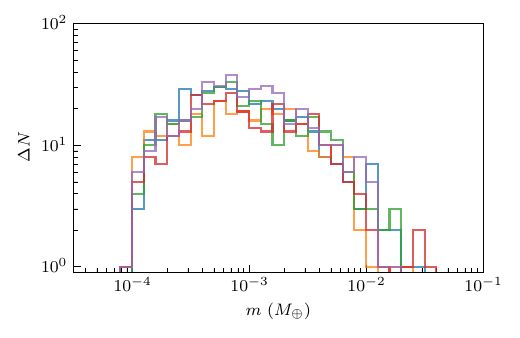}
	\caption{Differential mass distribution of the planetesimals formed near the initial disc substructure for the set of simulations with $\alpha=5\times10^{-4}$ drawn from the adopted initial mass function by \cite{gerbig2023}. There are 10 bins per decade in mass $m$ and $\Delta N$ is the number of planetesimals in each bin. Each colour corresponds to one of the five simulations.}
	\label{fig:IMF_ex}
\end{figure}

\cite{Lau2022} adopted the truncated power-law cumulative mass distribution from the fitting by \cite{abod2019}. However, the upper end of the distribution is not limited and \cite{Lau2024} noted that the largest planetesimal in the actual realisation depends on the total number of planetesimals. Therefore, here we adopted the Toomre-like instability criterion $Q_\mathrm{p}$ for the gravitational collapse of the dense filament induced by streaming instability and the initial mass function from \cite{gerbig2023}. The criterion for collapse is $Q_\mathrm{p}<1$ with
\begin{equation}
	Q_\mathrm{p}=\sqrt{\frac{\delta}{\mathrm{St}_\mathrm{avg}}}\frac{c_\mathrm{s}\Omega}{\pi G\Sigma_\mathrm{d, local}},
\end{equation}
and the mass-averaged Stokes number of the dust in the cell $\mathrm{St}_\mathrm{avg}$. The local dust surface density $\Sigma_\mathrm{d, local}$ was assumed to be 10 times of the averaged $\Sigma_\mathrm{d}$ of the cell. Motivated by the streaming instability simulations in \cite{Schreiber2018}, the small-scale diffusion parameter $\delta$ was set at $10^{-5}$. The model converts dust into planetesimals based on the prescription by \cite{DrazkowskaJ.2016} and \cite{Schoonenberg2018}. The $Q_\mathrm{p}$ criterion was combined with the smooth planetesimal formation activation function from \cite{Miller2021}, which is given by
\begin{equation}
	\mathcal{P_\mathrm{pf}}= \frac{1}{1+\exp{[10\times(Q_\mathrm{p}-0.75)]}}, \label{eq:Ppf}
\end{equation}
and evaluated at each radial grid cell. If any cell also satisfies the criterion of $\rho_\mathrm{d}/\rho_\mathrm{g}\ge1$, the local dust surface density for each dust species $i$ is reduced by
\begin{equation}
	\frac{\partial \Sigma_{\mathrm{d},i}}{\partial t}=-\mathcal{P_\mathrm{pf}}\Sigma_{\mathrm{d},i}\frac{\zeta}{t_{\mathrm{set},i}}.
\end{equation}
The planetesimal formation efficiency per settling time is $\zeta=10^{-3}$ and the settling time of dust species $i$ is $t_{\mathrm{set},i}\equiv1/(\mathrm{St}_i\Omega_\mathrm{K})$.

Then, the removed dust is summed over all dust species and added to the local planetesimal mass surface density. We first draw the location of a new planetesimal using the radial profile of the planetesimal mass surface density. Then, we draw the planetesimal mass according to the initial mass function given by \cite{gerbig2023}, which is resulting from the stability analysis of the dispersion relation for dust influenced by turbulent diffusion \citep{klahr2021}. The maximum and minimum masses associated with the unstable modes are given by
\begin{equation}
	m_{\max,\min} = \frac{9}{8} \sqrt{\frac{\pi}{2}}  \left( \frac{\delta}{\mathrm{St}_\mathrm{avg}}\right) ^{3/2}  \hat{h}_\mathrm{g}^3 \left(\frac{1}{Q_\mathrm{p}}\pm \sqrt{\frac{1}{Q_\mathrm{p}^2}-1}   \right) ^2 M_\odot,
\end{equation}
and the fastest-growing mode is given by
\begin{align}
	m_\mathrm{fgm} &= \frac{9}{8} \sqrt{\frac{\pi}{2}} \left( \frac{\delta}{\mathrm{St}_\mathrm{avg}}\right) ^{3/2} \hat{h}_\mathrm{g}^3 Q_\mathrm{p}^2 M_\odot\\
	&\approx 0.37\left( \frac{\delta/10^{-5}}{\mathrm{St}_\mathrm{avg}/0.1} \right)^{3/2} \left( \frac{\hat{h}_\mathrm{g}}{0.05} \right)^3 \left( \frac{Q_\mathrm{p}}{1} \right)^2  M_\mathrm{Ceres},
\end{align}
assuming isotropy in the small-scale diffusion. We refer the readers to Eq. (20) in \cite{gerbig2023} for the expression of the probability density function. Figure \ref{fig:IMF_ex} shows an example of the differential mass distribution of the planetesimals formed at about 6.5 au near the initial disc substructure for the set of simulations with $\alpha=5\times10^{-4}$.

As described in \cite{Lau2022}, the eccentricity $e$ and the inclination $i$ in radian are randomly drawn from two Rayleigh distributions with the scale parameters $10^{-6}$ and $5\times10^{-7}$ respectively. The rest of the angles of the orbital elements in radian are drawn randomly from 0 to $2\pi$. The physical radius $R_\mathrm{p}$ is calculated by assuming an internal density $\rho_\mathrm{s}$ of 1.5 $\mathrm{g}\ \mathrm{cm}^{-3}$. The drawn mass is subtracted from the surface density from the nearest radial grid cells and the realisation stops when the total remaining mass is less than the drawn mass. Any residue of the planetesimal mass surface density and the last drawn value of planetesimal mass is retained for the next time step to avoid bias towards the lower mass, that is, the drawn mass will be realised as soon as enough planetesimal surface density is available.

\subsection{Planetesimal evolution} \label{sec:plts}
The realised planetesimals, and a Solar-mass star, were then evolved by SyMBAp with full gravitational interactions as well as additional subroutines to include gas drag, the planet-disc interactions, pebble accretion, gas accretion and planetary gap opening. If two bodies collide, they were assumed to merge completely, that is, collisions are perfectly inelastic. At each communication step, on top of the newly formed planetesimals, the radial profiles of the disc were passed to SyMBAp. This included the gas components:
\begin{itemize}
	\item the gas surface density $\Sigma_\mathrm{g}$;
	\item the midplane temperature $T$;
	\item the gas disc scale height $H_\mathrm{g}$;
	\item the midplane gas density $\rho_\mathrm{g}$, and;
	\item the midplane pressure gradient parameter $\eta$,
\end{itemize}
and the dust component, for each dust species $i$,: 
\begin{itemize}
	\item the Stokes number St$_i$;
	\item the dust disc scale height $H_{\mathrm{d},i}$, and;
	\item the dust surface density $\Sigma_{\mathrm{d},i}$.
\end{itemize}
Also, the feedback to the disc was also passed to DustPy including
\begin{itemize}
	\item the dust mass loss due to pebble accretion (Sect. \ref{sec:pa});
	\item the gas mass loss due to gas accretion (Sect. \ref{sec:ga}), and;
	\item the change in gas surface density due to planetary gap opening (Sect. \ref{sec:method:K}),
\end{itemize}
which are further described in the respective subsections.

\subsubsection{Pebble accretion} \label{sec:pa}
The treatment for the pebble accretion rate of each planetesimal is identical to that presented in \cite{Lau2022} and summarised below. First, the pebble mass flux of dust species $i$ at any location is given by
\begin{equation}
\dot{M}_\mathrm{peb}=2\pi rv_{\mathrm{drift},i} \Sigma_{\mathrm{d},i}.
\end{equation}
The pebble drift speed of dust species $i$ is $v_{\mathrm{drift},i}=2\mathrm{St}_i|\eta|r\Omega_\mathrm{K}$ \citep{Weidenschilling1977}.
Then, we implemented the pebble accretion efficiency factor $\epsilon_{\mathrm{PA},i}$ by \cite{Liu2018} and \cite{Ormel2018} to calculate the fraction of the local pebble mass flux being accreted by each planetesimal or planet for dust species $i$. The pebble accretion rate by a planetesimal or a planet is then given by summing the contributions from all dust species, which is
\begin{equation}
	\dot{m}_\mathrm{pa}=\sum_i \epsilon_{\mathrm{PA},i} 2\pi r v_{\mathrm{drift},i}\Sigma_{\mathrm{d},i}.
\end{equation}
The mass of the accreted pebbles is then subtracted from the respective dust species and radial cell of the dust disc at the next immediate communication step. 

We did not implement the pebble isolation mass explicitly with a prescription \citep[e.g.][]{Lambrechts2014} in this work. Instead, as dust and gas evolve consistently in this model, the planetary gap opening by a planet (Sect. \ref{sec:method:K}) can interrupt the pebble flux capturing the process of pebble isolation within the model.

\subsubsection{Gas accretion} \label{sec:ga}
We followed \cite{Piso2014} and \cite{Bitsch2015} to prescribe the gas accretion rate with the modification by \cite{Chambers2021} to account for the energy released from pebble accretion. Gas accretion generally begins when the energy released from pebble accretion decreases enabling the cooling of the gas envelope. The gas accretion rate in this phase is
\begin{equation}
	\begin{split}
		\dot{m}_\mathrm{cool}=&\max\Bigg[ 0, 4.375\times10^{-9} \left(\frac{\kappa}{\mathrm{cm}^2\ \mathrm{g}^{-1}} \right)^{-1} \left( \frac{\rho_\mathrm{c}}{5.5\ \mathrm{g}\ \mathrm{cm}^{-3}}\right)^{-1/6} \times\\
		 & \left( \frac{m_\mathrm{c}}{M_\oplus} \right)^{11/3}\left( \frac{m_\mathrm{env}}{M_\oplus} \right)^{-1} \left( \frac{T}{81\mathrm{ K}}\right) ^{-1/2} M_\oplus \mathrm{yr}^{-1} - 15 \dot{m}_\mathrm{pa}\Bigg]
	\end{split}
\end{equation}
with the opacity of the gas envelope $\kappa$ and the density of the core $\rho_\mathrm{c}=5.5\ \mathrm{g}\ \mathrm{cm}^{-3}$. We note that the solid mass accreted from planetesimal accretion is negligible compared to pebble accretion in our simulations. We followed \cite{Brouwers2021} for the grain size near the Bondi radius of the gas envelope to calculate its Rosseland mean opacity. Assuming the Epstein regime, the incoming solids converge to the size
\begin{equation}
	R=\frac{\rho_\mathrm{g} v_\mathrm{th} v_\mathrm{frag}}{g_\mathrm{B}\rho_\bullet},
\end{equation}
with the midplane gas density $\rho_\mathrm{g}$, the thermal velocity $v_\mathrm{th}$, the gravity at the Bondi radius $g_\mathrm{B}$ and the monomer density $\rho_\bullet = 1.67\ \mathrm{g}\ \mathrm{cm}^{-3}$. The Rosseland mean opacity is given by
\begin{equation}
	\kappa=\frac{3Q_\mathrm{eff}\rho_\mathrm{d}}{4\rho_\bullet R\rho_\mathrm{g}}
\end{equation}
with the midplane dust density $\rho_\mathrm{d}$. The extinction efficiency is $Q_\mathrm{eff}=\min(0.6\pi R_\mathrm{s}/\lambda_\mathrm{peak},2)$ with the peak wavelength of the emission given by
\begin{equation}
	\lambda_\mathrm{peak}=\frac{0.29\ \mathrm{cm}}{T/K},
\end{equation}
where the temperature $T$ is assumed to be the local disc midplane temperature.

When the gas envelope $m_\mathrm{env}$ exceeds the solid core mass $m_\mathrm{c}$, gas accretion enters the runway phase following the treatment by \cite{Bitsch2015}. The gas accretion rate in this phase is determined by the gas stream flowing towards the planet \citep{Tanigawa2016}, which is
\begin{equation}
	\dot{m}_\mathrm{runaway}=0.29\Sigma_\mathrm{g}r^2\Omega_\mathrm{K} \left( \frac{m}{M_\odot}\right) ^{4/3}\hat{h}_\mathrm{g}^{-2}
\end{equation}
with the planet mass $m$. The accreted gas is then subtracted from the gas disc assuming the half-width of the accretion zone equals twice the Hill radius of the planet.

\subsubsection{Physical radius}
As the planetesimals or planets grow by many orders of magnitude in mass, the physical radius $R_\mathrm{p}$ is evaluated correspondingly. For mass $m$ less than 0.1 Earth mass, we assumed an internal density $\rho_\mathrm{s}$ of 1.5 $\mathrm{g}\ \mathrm{cm}^{-3}$, which is the same when the planetesimals are formed. For $m$ above 0.1 $M_\oplus$ but less than 5 $M_\oplus$, we followed \cite{Seager2007} for the mass-radius relationship of rocky planets, which is
\begin{equation}
	\log\left( \frac{R_\mathrm{p}}{3.3R_\oplus} \right) = -0.209 + \frac{1}{3}\log\left( \frac{m}{5.5M_\oplus} \right)-0.08\left( \frac{m}{5.5M_\oplus} \right)^{0.4}
\end{equation}
with the radius of Earth $R_\oplus$.
For $m$ above 5 $M_\oplus$, we followed the mass-radius relationship applied in \cite{Matsumura2017}, which is
\begin{equation}
	R_\mathrm{p}=1.65\sqrt{\frac{m}{5M_\oplus}}R_\oplus.
\end{equation}

\subsection{Planetary gap opening}\label{sec:method:K}
For all planetesimals and planets, the non-dimensional gap opening factor \citep{Kanagawa2015} was evaluated, which is given by
\begin{equation}
K=\left( \frac{m}{M_\odot} \right) ^2\hat{h}_\mathrm{g}^{-5}\alpha^{-1}.
\end{equation}
Since the treatment in Sect. \ref{sec:gap} to attain a target gas profile does not change the actual turbulence experienced by the dust and we only consider the torque exerted on the gas disc, planetary gap opening was applied through the modified $\alpha'$-parameter. When $K>0.25$, we impose a planetary gap to the gas disc by dividing $\alpha'$ by the ratio of the perturbed surface density to the unperturbed one $\Sigma_\mathrm{g}/\Sigma_{\mathrm{g},0}$. In other words, the planetary gap is imposed when the change caused by the corresponding body is more than about 1\% of the unperturbed surface gas density.

The effects on $\alpha'$ were multiplied when more than one planet can open a gap, that is, for all gap opening planets $i$ 
\begin{equation}
\alpha' = \frac{\alpha}{F \cdot \prod_{i} \left(\Sigma_\mathrm{g}/\Sigma_{\mathrm{g},0}\right)_i},
\end{equation}
with the function to impose the initial disc gap $F$ (see Eq. (\ref{eq:F_gap})).

We note that some small but short-period chaotic movements of massive planets can prevent the disc from converging to a quasi-steady state and the speed of code is significantly reduced. Therefore, we allowed $\alpha'(r)$ to reach the target value exponentially on a relaxation timescale of $250\times(10^{-3}/\alpha)$ years, which is below the viscous evolution timescales of the disc in the adopted radial domain of the simulations.

We adopted the empirical formula by \cite{Duffell2020} for the gap profile, which is
\begin{equation}\label{eq:duffell}
	\frac{\Sigma_\mathrm{g}}{\Sigma_{\mathrm{g},0}} = \left(1+\frac{0.45}{3\pi}\frac{\tilde{q}^2(r)}{\alpha\hat{h}_\mathrm{g}^5}\delta(\tilde{q}(r))\right)^{-1}
\end{equation}
with $\hat{h}_\mathrm{g}$ evaluated at the planet's location. The radial profile function $\tilde{q}(r)$ is defined by
\begin{equation}
	\tilde{q}(r)\equiv\frac{q}{\left\lbrace 1+D^3\left[ (r/r_\mathrm{p})^{1/6}-1 \right] ^6 \right\rbrace^{1/3} }
\end{equation}
with the mass ratio $q\equiv m/M_\odot$, the planet's radial distance from the star $r_\mathrm{p}$ and the scaling factor $D\equiv 7\hat{h}_\mathrm{g}^{-3/2}\alpha^{-1/4}$.
The function $\delta(x)$ is given by
\begin{equation}\label{eq:del_x}
\delta(x) = \begin{cases}
		1+(x/q_w)^3, 	 &\mathrm{if}\ x<q_\mathrm{NL}\\
		\sqrt{q_\mathrm{NL}/x}+(x/q_w)^3, &\mathrm{if}\ x\geq q_\mathrm{NL}
	\end{cases}
\end{equation}
with the factor $q_\mathrm{NL}=1.04\hat{h}_\mathrm{g}^3$ and the factor $q_w=34 q_\mathrm{NL}\sqrt{\alpha/\hat{h}_\mathrm{g}}$. We note that there is a minor discrepancy in the form of $\delta(x)$ for the case of $x<q_\mathrm{NL}$ between the text and the Python code presented in \cite{Duffell2020}. We confirm that the expression given above by Eq. (\ref{eq:del_x}) is consistent with the mentioned Python code, which is continuous and more appropriate \citetext{private communication, Duffell, 2022}.

\subsection{Gas drag and planet-disc interactions}\label{sec:mig}
All bodies experience the combined effects of aerodynamic gas drag and the planet-disc interactions. For low-mass planet without gap opening, the treatment for gas drag and planet-disc interaction are identical to that presented in \cite{Lau2022} and summarised below.

We adopt the aerodynamic gas drag by \cite{Adachi1976}, which is
\begin{equation}
	\mbox{\boldmath $a$}_\mathrm{drag}=-\left( \frac{3C_\mathrm{D}\rho}{8R_\mathrm{p}\rho_\mathrm{s}}\right) v_\mathrm{rel}\mbox{\boldmath $v$}_\mathrm{rel}
\end{equation}
with the drag coefficient $C_\mathrm{D}$ and, the relative velocity between the planetesimal and the gas $\mbox{\boldmath $v$}_\mathrm{rel}$. The gas flow is assumed to be laminar and cylindrical, where the magnitude is given by $r\Omega_\mathrm{K}(1-|\eta|)$. As the planetesimals in this work are well larger than a kilometre in size, the large Reynolds number case is generally applicable, that is, $C_\mathrm{D}=0.5$ \citep{Whipple1972}. The gas density $\rho$ at the planetesimal's position $z$ above the midplane is given by $\rho=\rho_\mathrm{g}\exp(-0.5z^2/H_\mathrm{g}^2)$.

For type-I damping and migration, we adopted the prescription based on dynamical friction by \cite{Ida2020}. The timescales for the isothermal case and finite $i$, while $i<\hat{h}_\mathrm{g}$, (Appendix D of \cite{Ida2020}) were implemented. The evolution timescales of semimajor axis, eccentricity and inclination are defined, respectively, by
\begin{equation}
	\tau_a\equiv-\frac{a}{\mathrm{d}a/\mathrm{d}t},\tau_e\equiv-\frac{e}{\mathrm{d}e/\mathrm{d}t}, \tau_i\equiv-\frac{i}{\mathrm{d}i/\mathrm{d}t}.
\end{equation}
These timescales are given by, with $\hat{e}\equiv e/\hat{h}_\mathrm{g}$ and $\hat{i}\equiv i/\hat{h}_\mathrm{g}$,
\begin{equation}
	\tau_a = \frac{t_\mathrm{wav}}{C_\mathrm{T}\hat{h}_\mathrm{g}^2}\left[ 1+\frac{C_\mathrm{T}}{C_\mathrm{M}}\sqrt{\hat{e}^2+\hat{i}^2}\right], \label{eq:tau_a}
\end{equation}
\begin{equation}
	\tau_e =1.282t_\mathrm{wav}\left[ 1+ \frac{(\hat{e}^2+\hat{i}^2)^{3/2}}{15} \right],
\end{equation}
\begin{equation}
	\tau_i =1.838t_\mathrm{wav}\left[ 1+ \frac{(\hat{e}^2+\hat{i}^2)^{3/2}}{21.5} \right].
\end{equation}
The characteristic time $t_\mathrm{wav}$ \citep{Tanaka2002} is given by
\begin{equation}
	t_\mathrm{wav}=\left( \frac{M_\odot}{m}\right) \left( \frac{M_\odot}{\Sigma_{\mathrm{g}}r^2}\right) \left( \frac{\hat{h}_\mathrm{g}^4}{\Omega_\mathrm{K}}\right),\label{eq:twav}
\end{equation}
where $\Sigma_{\mathrm{g}}$ and $\hat{h}_\mathrm{g}$ are retrieved from the local radial cell of the disc model. The normalised torques $C_\mathrm{M}$ and $C_\mathrm{T}$ are given by
\begin{equation}
	C_\mathrm{M}=6(2p_\Sigma-q_T+2),  \label{C_M}
\end{equation}
\begin{equation}
	C_\mathrm{T}=2.73+1.08p_\Sigma+0.87q_T,\label{C_T}
\end{equation}
with $p_\Sigma\equiv-\mathrm{d}\ln\Sigma_{\mathrm{g}}/\mathrm{d}\ln r$ and $q_T\equiv-\mathrm{d}\ln T/\mathrm{d}\ln r$. The three timescales were then applied to the equation of motion
\begin{equation}
	\mbox{\boldmath $a$}=-\frac{v_\mathrm{K}}{2\tau_a}\mbox{\boldmath $e$}_\theta-\frac{v_r}{\tau_e}\mbox{\boldmath $e$}_r-\frac{v_\theta-v_\mathrm{K}}{\tau_e}\mbox{\boldmath $e$}_\theta-\frac{v_z}{\tau_i}\mbox{\boldmath $e$}_z
\end{equation}
in the cylindrical coordinates $(r,\theta,z)$ that the velocity of the embryo $\mbox{\boldmath $v$}=(v_r,v_\theta,v_z)$. And, the local Keplerian velocity $v_\mathrm{K}$ was evaluated at the instantaneous $r$ of the particle.

As the planet grows and opens a gap in the disc, \cite{Kanagawa2018} suggested that the magnitude of the torque scales linearly with the local surface density providing a smooth transition to the high-mass (type-II) regime of planet migration. Since the dependence of $t_\mathrm{wav}$ on $\Sigma_{\mathrm{g}}$ as in Eq. (\ref{eq:twav}) is retrieved from the local grid cell, the above treatment combined with gap opening (Sect. \ref{sec:method:K}) can already capture this transition.

We note that planetary gap opening (Sect. \ref{sec:method:K}) and planet migration are the results of planet-disc interactions, which are physically coupled by the action-reaction pair. However, we have adopted two independent prescriptions for each of them since a prescription for a torque profile which is suitable for a one-dimensional model and a general planet mass is still missing. Further discussions on the adopted treatments are in Sect. \ref{sec:torque_prof}.

\subsection{Numerical setup}\label{sec:num}
The time step for SyMBAp $\tau=0.2$ yr was used and particles were removed if the heliocentric distance is less than 4 au or greater than $100$ au. The additional subroutines for the evolution of the $N$-body particles were added to SyMBAp as
\begin{equation}
	\mathcal{P}^{\tau/2}\mathcal{M}^{\tau/2}\mathcal{N}^{\tau}\mathcal{M}^{\tau/2}\mathcal{P}^{\tau/2}.
\end{equation}
The operator $\mathcal{P}$ handles the effect of pebble accretion, gas accretion and gap opening, $\mathcal{M}$ handles the effect of gas drag and planet-disc interactions, and $\mathcal{N}$ is the second-order symplectic integrator in the original SyMBAp. The operators $\mathcal{P}$ and $\mathcal{M}$ operate in the heliocentric coordinates and $\mathcal{N}$ operates in the democratic heliocentric coordinates so coordinate transformation is required at each step.

Since disc dissipation is not included in this work, all simulations stop at 2 Myr, which is the typical timescale that internal photoevaporation becomes significant to the disc \citep[e.g.][]{Owen2010,Owen2011,Picogna2019,Garate2021}. We note a numerical difficulty when multiple giant planets are produced, which causes multiple deep planetary gaps in the disc and a small integration time step is required for the disc. Each simulation requires a wall-clock time of 2 to 4 weeks. We tested two values of $\alpha=\{3,5\}\times10^{-4}$ and five simulations were conducted for each to evaluate the statistical effect.

\section{Results} \label{sec:results}
\subsection{The case of $\alpha=5\times10^{-4}$} \label{sec:al54}
\begin{figure}
	\centering
	\includegraphics{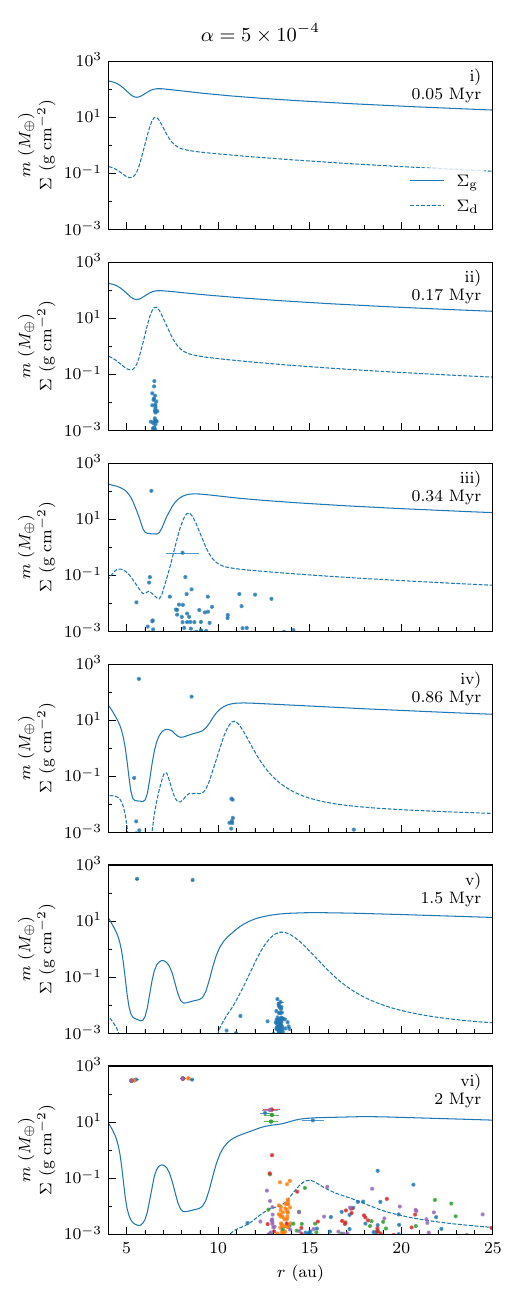}
	\caption{Six key timestamps demonstrating sequential planet formation in one of the simulations with $\alpha=5\times10^{-4}$. Each panel shows the radial profiles of the gas surface density $\Sigma_\mathrm{g}$ (solid line), the dust surface density $\Sigma_\mathrm{d}$ (dashed line) and, the mass $m$ and the semimajor axis $r$ of the massive bodies (dot) at the noted time. The final panel also shows the massive bodies from the rest of the simulation set with each colour showing one of the five simulations.}
	\label{fig:seq_pl_al54}
\end{figure}
\begin{figure}
	\centering
	\includegraphics{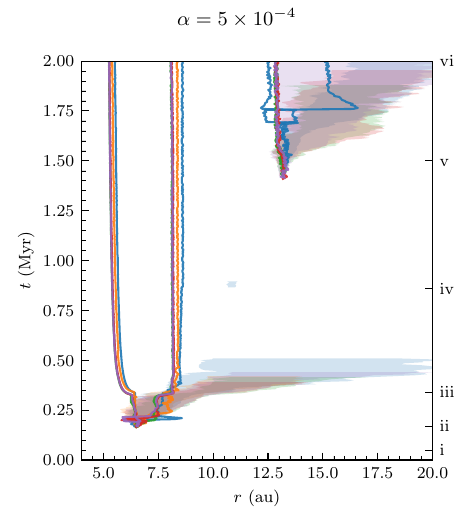}
	\caption{Tracks of the massive bodies in the simulation with $\alpha=5\times10^{-4}$. Each colour shows one of the five simulations corresponding to those in the final panel of Fig. \ref{fig:seq_pl_al54}, where the key timestamps are also denoted along the time axis on the right. The solid lines show the semimajor axis of the bodies reached 10 $M_\oplus$ by the end of the simulations. The shaded areas indicate the extents of the lower and upper quartiles of the semimajor axes of all bodies when the total number is above 50.}
	\label{fig:track_al54}
\end{figure}

\subsubsection{Formation and evolution of massive bodies} \label{sec:al54_archi}
Figure \ref{fig:seq_pl_al54} presents one of the simulations with $\alpha=5\times10^{-4}$ with the panels showing the six key timestamps. The solid and dashed lines show the profiles of the gas surface density $\Sigma_\mathrm{g}$ and the dust surface density $\Sigma_\mathrm{d}$ respectively. The dots show the mass $m$ and the semimajor axis $r$ of the massive bodies, with the error bar indicating the extent of the apoapsis and periapsis for those above 10 $M_\oplus$. The final panel (vi), which presents the end results at 2 Myr, also includes the massive bodies from the rest of the simulation set with each colour showing one of the five simulations.

Figure \ref{fig:track_al54} presents the evolution of the semimajor axis of the massive bodies with each colour showing one of the five simulations corresponding to those in the final panel of Fig. \ref{fig:seq_pl_al54}. The solid lines show the ones reaching more than 10 $M_\oplus$ by the end of the simulations. The shaded areas indicate the extents of the lower and upper quartiles of semimajor axes, which is only shown when the total number of bodies is above 50 for a meaningful representation.

At 0.05 Myr (i), the imposed initial substructure has reached the target shape and dust has started to accumulate. At 0.17 Myr (ii), the midplane volumetric dust-to-gas ratio of the disc at about 6.5 au reaches the criteria and planetesimal formation starts. Since these bodies are naturally born in a dust-rich environment, where the dust surface density is more than an order of magnitude higher than the unperturbed case, they can grow rapidly by pebble accretion. The core has also migrated towards the migration trap, which is slightly interior to the peak of the pressure bump, but not further inside as shown by the track in Fig. \ref{fig:track_al54}.

At 0.34 Myr (iii), the first massive core has entered the runaway gas accretion phase and opened a significant gap in the disc as it becomes a gas giant ($>100M_\oplus$). The less massive core and planetesimals, which are also formed from the initial pressure bump, are being scattered mainly to wider orbits as shown by the tracks in Fig. \ref{fig:track_al54}. While most planetesimals have been scattered out of the system, the orbit of the second-most massive planetary core is circularised near 8 to 9 au, and it continues to grow.

At 0.86 Myr (iv), the second core also starts runaway gas accretion but at a much later time relative to the first one. When the second gas giant opens a gap in the disc, the dust near its location follows the sudden change in the gas profile and is pushed away from the forming gas giant to both inner and outer part of the disc. This corresponds to the formation of a small batch of planetesimals near 11 au shown in Fig. \ref{fig:track_al54}.

At 1.5 Myr (v), the second gas giant also reached approximately one Jupiter mass with another planetary gap fully opened. A new pressure bump is steadily formed at the outer edge of this gap and dust re-accumulates at about 13 to 14 au, which contains a part of the leftover dust from the initial dust trap and the dust drifted from the outer disc. Another generation of planetesimals is formed at this location. A minor instability occurred between two newly formed massive cores at around 1.75 Myr that widens their radial separation.

Due to the late formation of the second generation of planetary cores, which ultimately form a pair of ice giants ($10-100 M_\oplus$), they remain in the thermal contraction phase of gas accretion at the end of the simulation at 2 Myr (vi). A compact chain of giant planets is produced spanning from 5 to 15 au, with a pair of gas giants formed from the initial pressure bump and a pair of ice giants formed over 1 Myr later from the edge of the planetary gap opened by the outer gas giant. The orbital periods of the inner pair are in near 2:1 commensurability and those of the outer pair are in near 4:3 commensurability.

Across the five random simulations, the final panel of Fig. \ref{fig:seq_pl_al54} and Fig. \ref{fig:track_al54} show very similar results for the gas giant pair formed in the first generation. For the next generation of planet formation, further stochasticity presents. Two (blue and green) simulations produce a pair of ice giants and another two (red and purple) produce only one ice giant. One simulation (orange) shows no ice giant at 2 Myr but a swarm of still-growing planetesimals and planet embryos.

\subsubsection{Dust mass budget} \label{sec:al54_dust}
\begin{figure}
	\centering
	\includegraphics{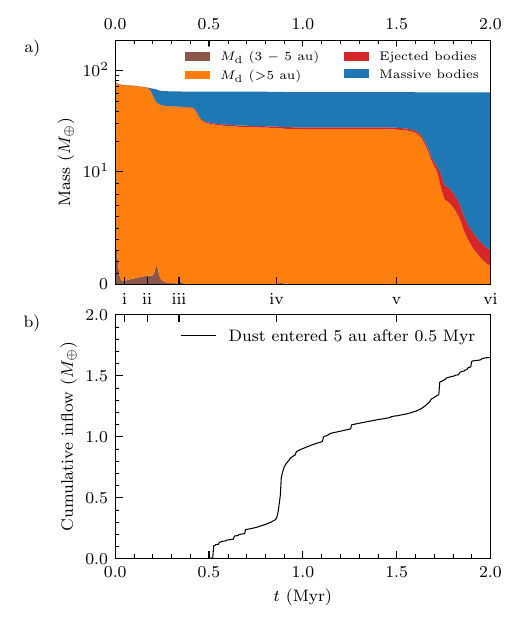}
	\caption{Evolution of the solid mass in the simulation with $\alpha=5\times10^{-4}$ that corresponds to the one presented by the colour blue in Fig. \ref{fig:seq_pl_al54} \& \ref{fig:track_al54}. The six timestamps in Fig. \ref{fig:seq_pl_al54} are also denoted on the time axis. a) Solid mass budget. The solid mass is divided into four categories: the dust mass $M_\mathrm{d}$ inside and outside of 5 au, solids bound in massive bodies and ejected massive bodies. It shows a high planet formation efficiency that the majority of solid mass (85\% of the initial dust mass beyond 5 au) are converted into massive bodies.
	b) Cumulative inflow of dust entered 5 au after 0.5 Myr. A total of about 1.6 $M_\oplus$ of inflow to the inner disc is recorded over the next 1.5 Myr up to the end of the simulation.}
	\label{fig:dust_t}
\end{figure}

Figure \ref{fig:dust_t}a shows the solid mass budget throughout the simulation presented by the colour blue in Fig. \ref{fig:seq_pl_al54} \& \ref{fig:track_al54}. The solid mass is divided into four categories, which are the dust mass $M_\mathrm{d}$ inside and outside of 5 au, solids bound in massive bodies, and massive bodies ejected out of the simulation domain. The six timestamps in Fig. \ref{fig:seq_pl_al54} are also denoted here on the time axis.

After the initial substructure is imposed, the dust mass between 3 au and 5 au (inner disc) decreases sharply due to the inward drift of dust, while the dust supply from 5 au and beyond (outer disc) is stopped at the pressure bump. At 0.05 Myr (i), the pressure bump is saturated with dust and the dust mass in the inner disc increases due to leakage by turbulent diffusion.
Planetesimals start to form and grow by pebble accretion at 0.17 Myr (ii), which start converting dust into massive bodies. The conversion is paused when the first core starts to perturb the disc and stops pebble accretion.

Shortly after 0.2 Myr, there is a spike in the dust mass inside of 5 au. This is caused by the gap opening in a dust rich location as the first planetary core reaches the mass of $10 \ M_\oplus$. Similarly, another spike occurs at about 0.34 Myr (iii) when the first core enters the runaway gas accretion phase but it is much smaller as dust is already depleted around the planet. Pebble accretion resumes for the second core at about 0.4 Myr after its orbit has been circularised, which also causes the change in the dust mass.

Planetesimal formation occurs again at 1.5 Myr (v) and pebble accretion continues to convert the remaining dust to massive bodies. The final (vi) masses of the four categories show that the majority of solids, or 85\% of the initial dust mass beyond 5 au, are eventually incorporated into massive bodies.

Figure \ref{fig:dust_t}b shows the cumulative inflow of dust that crossed 5 au after 0.5 Myr, which is the time when the first gas giant has reached approximately one Jupiter mass. The subsequent dust inflow to the inner Solar System is, on average, 0.5 to 1 $M_\oplus\mathrm{ Myr}^{-1}$ or, in total, about 1.6 $M_\oplus$ including two significant episodic inflows to the inner disc.
 
The first one occurs shortly at about 0.86 Myr (iv) as the second planetary core enters the runaway gas accretion phase and opens a gap in the disc. The dust near its location follows the sudden change in the gas disc and is pushed by the forming gas giant to both inner and outer part of the disc, which is also shown in the profiles of the surface densities (Fig. \ref{fig:seq_pl_al54} iv). The second one corresponds to a small instability occurs between the two newly formed massive cores at around 1.75 Myr and perturbs the disc (Fig. \ref{fig:track_al54}).

\subsection{The case of $\alpha=3\times10^{-4}$} \label{sec:al34}
\begin{figure}
	\centering
	\includegraphics{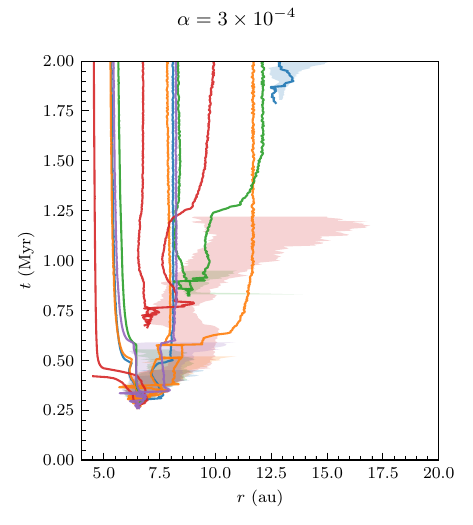}
	\caption{Tracks of the massive bodies in the simulations with $\alpha=3\times10^{-4}$ presented in the same manner as in Fig. \ref{fig:track_al54}.}
	\label{fig:track_al34}
\end{figure}
\begin{figure}
	\centering
	\includegraphics{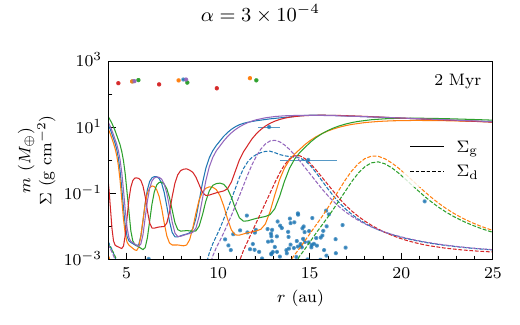}
	\caption{Results of the set of five simulations with $\alpha=3\times10^{-4}$. Similar to the final panel of Fig. \ref{fig:seq_pl_al54} except the surface density profiles are shown for all simulations.}
	\label{fig:seq_pl_al34}
\end{figure}

Figure \ref{fig:track_al34} presents the tracks of the massive bodies for the set of simulations with $\alpha=3\times10^{-4}$ in the manner of Fig. \ref{fig:track_al54}. Figure \ref{fig:seq_pl_al34} presents the end results to the final panel of Fig. \ref{fig:seq_pl_al54}. The radial profiles of the surface densities are also shown for the whole set of simulations. Compared to the case of $\alpha=5\times10^{-4}$ in Sect. \ref{sec:al54}, a larger variation across the simulations is shown. For all simulations, planetesimal formation occurs at about 0.25 Myr, which is about 0.1 Myr later than the case of $\alpha=5\times10^{-4}$.

In the simulations denoted by the colour red in Fig. \ref{fig:track_al34}, a massive core is scattered through the migration trap by another core and is lost to the inner simulation boundary. Later at about 0.6 Myr, the next generation of planetesimals are formed resulting in two massive cores. Similarly, in the simulations denoted by the colour green, only one core is formed from the initial pressure bump and two is formed from the subsequent generation.

In the simulations denoted by the colour blue and purple, two cores are formed from the initial bump. The second generation of planet formation occurs at about 1.75 Myr for the former one, while only a concentrated dust ring presents at the outer edge of the planetary gap for the latter one at the end of the simulation. And, the simulation denoted by the colour orange forms three gas giants from the initial pressure bump and no further planet formation occurs before the simulation ends.

Figure \ref{fig:seq_pl_al34} summarises the final architecture of the planets where three out of the five simulations form three gas giants and the remaining two (blue and purple) form two. Also, a significant dust ring remains respectively for all simulations external to their outermost gas giant.

\section{Discussions} \label{sec:dis}
\subsection{Sequential planet formation} \label{sec:seq_pl}
The results presented in Sect. \ref{sec:results} demonstrate a complete scenario of sequential planet formation. In the case of $\alpha=5\times10^{-4}$ (Sect. \ref{sec:al54}), two gas giants are formed from the initial disc substructure. As the outer gas giant has reached its final mass and a steady planetary gap is opened, dust re-accumulates at a later time near the new pressure bump triggering the next generation of planet formation. The case of $\alpha=3\times10^{-4}$ shows similar trend despite of the greater degree of stochasticity.

Comparing the results with the `inside-out' planet formation scenario by \cite{chatterjee2013} for Kepler systems, we note that gas giant that has reached its final mass is more likely to trigger the next generation of planet formation. For low-mass planets that cannot open a significant gap in the disc ($\lesssim100M_\oplus$), dust leakage from the pressure bump at the outer edge of the gap is significant and requires a large supply from the outer disc to reach the conditions for planetesimal formation. Even if planetesimals may form, they are under greater gravitational influence of the planets as the width of the gap scales with $m^{1/2}$ \citep{kanagawa2016,Duffell2020} while the Hill radius scales with $m^{1/3}$. In this case, the perturbation from the planet is more likely to prevent the growth of the planetesimals as pebble accretion, particularly when the planet continues to grow, is sensitive to the relative velocity as noted in \cite{Lau2022}. These planetesimals will likely be scattered out of the system as well, if the planet enters the runaway gas accretion phase to become a gas giant. Therefore, the outer edge of the planetary gap opened by a steady gas giant is a much more favourable environment for the next generation of planet formation.

\subsubsection{Architecture of the resulting systems} \label{sec:div}
In the case of $\alpha=5\times10^{-4}$ (Sect. \ref{sec:al54}), the delay in the formation of the second generation of planets directly shortens the time available for their growth. Therefore, they remain at about $10M_\oplus$ by the end of the simulations. Although disc dissipation is not included in the current model, the sequential planet formation scenario provides the delay in formation time of the ice giants required by the models explaining their masses \citep[e.g.][]{lee2014,Ogihara2020,Raorane2024}. At the end of the simulations, the second generation planetesimals and embryos formed still remain in the system as the ice giants are not able to scatter them. This results in a system with diversity that consists of gas giants, ice giant(s) and small massive bodies.

Since the radial separation of the two generations of planets is determined by the gap width, the resulting system is compact with the four giants planets from 5 to 15 au. The gas giants formed from the initial substructure are also commonly in the 2:1 mean-motion resonance. While a further test with a larger sample size is required, this may justify the compact chain of giant planets adopted in the initial conditions of the Nice model \citep[e.g.][]{Tsiganis2005,morbidelli2005} and the early instability model \citep[e.g.][]{Clement2017,Deienno2018} from the formation point of view.

Due to the computational cost, only two values of $\alpha$ are tested in this work where a larger degree of stochasticity is presented for the case of $\alpha=3\times10^{-4}$ (Sect. \ref{sec:al34}). Other than $\alpha$, which determines the evolution timescale of the disc, the resulting planetary system should also be sensitive to the initial disc mass $M_\mathrm{disc}$, the characteristic radius $r_\mathrm{c}$ and the location of the initial disc substructure. Upon the availability of TriPoD (Pfeil et al. submitted), which is a simplified three-parameter dust coagulation model, a more extensive parameter study shall become computationally feasible to study the diversity of planetary systems.

\subsubsection{Dust mass budget} \label{sec:pl_eff}
The resulting solid mass budget (Fig. \ref{fig:dust_t}a) shows a high planet formation efficiency for the presented simulation with $\alpha=5\times10^{-4}$. The common gap opened by the two gas giants is very effective in preventing dust from drifting through. As a result, the remaining dust is retained in the disc for a prolonged period of time and preserved solids for the subsequent planet formation.

After the formation of the first gas giant, the quasi-steady inflow and the episodic inflows result in a total of about 1.6 $M_\oplus$ of outer disc dust flowing into the inner disc over a time period of 1.5 Myr (Fig. \ref{fig:dust_t}b). The leftover planetesimals formed from the initial disc substructure are also generally scattered outward by the rapid formation of the first gas giant (Fig. \ref{fig:track_al54}). This indicates the chemical division caused by the first gas giant is robust. While further tests with the correct masses of the giant planets and their time of formation are required, this formation scenario may provide the required rapid formation of Jupiter's core to prevent significant exchange of the non-carbonaceous and carbonaceous reservoirs in the early Solar System \citep{Kruijer2017}.

\cite{Stammler2023} studied the efficiency of dust trapping by planetary gaps corresponding to different planetary masses and values of $\alpha$. Their results show a dust leakage rate of about $1 M_\oplus\mathrm{Myr}^{-1}$ for a gap created by a Saturn-mass planet with $\alpha=10^{-4}$. This is broadly consistent with the presented leakage rate of about $0.5 M_\oplus\mathrm{ Myr}^{-1}$ with the episodic inflows due to the dynamical instabilities excluded, since the planetary gap prescription used here is about 30\% deeper than the one given by \cite{kanagawa2016} (see Fig. 8 of \cite{Duffell2020}) and the planet is about three times more massive in our case.

The small inflow of dust from the outer disc is likely to be preferentially accreted by a proto-Earth and proto-Venus as pebble accretion is more efficient for bodies with higher mass and dynamically colder orbits. This may explain Earth's chemical abundances relative to Mars and Vesta \citep{kleine2023} without causing a significant growth to form super-Earth. This result also suggests against the scenario of significant growth by pebble accretion in the inner Solar System \citep[e.g.][]{johansen2021}.

\subsection{Comparison with the Solar System} \label{sec:sol}
We note that the results of $\alpha=5\times10^{-4}$ (Sect. \ref{sec:al54}) show a pair of Jupiter-mass giants, instead of one Jupiter- and one Saturn-mass giants. Although the second gas giant entered the runway gas accretion about 1 Myr later than the first one, there is no significant mass difference in the end. This is likely due to the absence of disc dissipation, which results in a high gas surface density throughout the simulation. Further developments of the model to include photoevaporation are required to provide a complete scenario of the formation of the early outer Solar System and planet formation in general. We anticipate that the remaining dust in the protoplanetary disc will form a belt of planetesimals resembling the scenario proposed by \cite{carrera2017}. In this case, these planetesimals will remain dynamically cold while growth by pebble accretion is prohibited due to the depletion of gas and dust. After disc dissipation, the $N$-body part (SyMBAp) can continue to model the long-term evolution of the system and show if a Nice model-like instability can occur among the giant planets.

The formation of the gas giants in the results is likely too quick compared to the meteoritic record \citep{Kruijer2017,Kruijer2020}. We note that the composition and the opacity of the envelope of gas giant, which are critical to the gas accretion rate, are still an active field of research \citep[e.g.][]{Szulagyi2016,lambrechts2019,schulik2019,ormel2021}. And, different gas accretion prescriptions are adopted among recent planet formation models \citep[e.g.][]{liu2019,Bitsch2019,matsumura2021,Chambers2021,Lau2024}. Further investigations on the different recipes and their consequences are required to match the formation history of the Solar System's giant planets.

While the source of the initial disc substructure is not investigated in this work, the water ice line in the early Solar System has been proposed as a key feature in reproducing the Solar System by multiple works \citep[e.g.][]{Morbidelli2016,Morbidelli2022,Brasser2020,Charnoz2021,Chambers2023}. Further investigations are required to determine the criteria at the water ice line in the early Solar System to trigger planet formation, particularly, the change in the surface density required.

\subsection{Other recent works} \label{sec:lit}
\subsubsection{Predictions of planet formation at pressure bumps}
\cite{Xu2024} made theoretical predictions on the architecture of the planetary systems assuming efficient planet formation at pressure bumps. They concluded three main pathways: slow core formation, fast core formation but slow gas accretion and, fast core formation and gas accretion.

While \cite{Lau2022} and \cite{Jiang2023} show that the high dust concentration at a pressure bump will likely favour rapid growth of core, the slow core formation can be possible if the planetesimals formed are dynamically heated to an extent that pebble accretion is halted but they still remain close to there birthplace.

The case of fast core formation but slow gas accretion is suggested to form a chain of super-Earths or potentially Saturn-mass planets over a prolonged period of time. This case is similar to the scenario proposed by \cite{chatterjee2013}. As discussed in Sect. \ref{sec:seq_pl}, the planetary gap formed by super-Earths may not be able to trap a significant amount of dust and trigger planetesimal formation. And, even if planetesimals could form, they are likely much closer to the core and cannot grow efficiently by pebble accretion.

The case of fast core formation and fast gas accretion is suggested to form a chain of gas giants. This prediction is the closest to the results shown in Sect. \ref{sec:results} while two to three gas giants can form from the pressure bump in the presented work. We note that the number of cores that can form from each pressure bump likely depends on the amplitude and the width of the dust trap, while this has not been tested in the presented work. This requires an extensive parameter study and more random simulations per set of parameters.

Although dust trap favours planetesimal formation, we emphasise that insitu planet formation at pressure bumps is unlikely a general solution to the diversity of exoplanetary systems. In particular, for the observed compact planet chains in resonance, a more probable formation scenario is that the cores are formed at a temporary pressure bump which migrate subsequently through the disc. Multiple works have studied the scenario that the inner most planet can be trapped at the disc edge \citep[e.g.][]{terquem2007,cossou2013,brasser2018,huang2023} and the subsequent inward-migrating planets can form a resonant chain through convergent inward migration \citep[e.g.][]{tamayo2017,delisle2017,macdonald2018,wong2024}.

\subsubsection{Sandwiched planet formation}
With gas and dust hydrodynamics simulations, \cite{pritchard2024} proposed the `sandwiched planet formation' scenario where planet formation can occur with the dust trapped between two massive planets that each creates a pressure maximum. The authors already noted that formation of the planets and dust fragmentation are not modelled, which may have critical effect to the dust concentration between the planets. Furthermore, from the results  presented above (Sect. \ref{sec:results}), we also note that if planetesimals could form between the planets, they are likely under a great gravitational influence from the massive planets. This will prevent them from growing efficiently by accreting pebbles, or, more likely, scatter them. Nonetheless, their work confirms dust rings can be created by massive planets with hydrodynamics simulations and these are preferred locations of planet formation.

\subsection{Caveats}
\subsubsection{Initial disc substructure}
In this work, we studied the consequence of a substructure in the disc that can trigger planetesimal formation, with the location motivated by that of Jupiter. Although multiple non-planetary mechanisms are proposed in the field as discussed in Sect. \ref{sec:intro}, the parameter space, including the location, amplitude, width and lifetime, is not explored in this work and requires future investigations. For instance, from some test runs, we note that the amplitude of the substructure needs to be large enough to trap dust effectively and trigger planetesimal formation, although non-axisymmetric features that may aid dust concentration such as vortices \citep[e.g.][]{Barge1995,Tanga1996} are not considered. Sequential planet formation also cannot occur if the dust mass remaining is not enough to trigger the next generation of planet formation. We emphasise that the criteria to form planetesimals are not trivial to satisfy in typical disc conditions. This work only focuses on a case where planetesimal formation is possible. However, by combining the unknowns in the shape and location of the initial disc substructure with different disc parameters, we expect the model can produce a variety of planetary systems through a parameter study, where sequential planet formation may not always occur.

\subsubsection{Planet migration and gap opening} \label{sec:torque_prof}
At the end of Sect. \ref{sec:mig}, we note that planetary gap opening and planet migration are treated by independent prescriptions while they are both the results of planet-disc interactions and are physically coupled. Also, multiple works \citep[e.g.][]{Lin1986,Armitage2002,DAngelo2010} have studied both effects consistently and provided formulae for the torque density profile exerted by a planet on the disc. However, upon applying the formula given by \cite{DAngelo2010} in our 1-D model, we note that the gap opened by a Jupiter-mass planet is much narrower than that described in \cite{Duffell2020}, which is also given by Eq. (\ref{eq:duffell}). Since this gap profile is tested against a set of 2-D hydrodynamical simulations in a general parameter space and is consistent with other works \citep[e.g.][]{Kanagawa2015}, we have opted to prescribe the gap profile and planet migration separately. We also note that a fixed temperature profile is adopted in this work, which implies that the effect of shock heating \citep[e.g.][]{Zhu2015,Rafikov2016} is neglected while its effect is likely more significant in the outer disc where less irradiation from the star is received. Nonetheless, upon the availability of a general torque formula applicable to a 1-D model, this part of the model shall be modified for consistency, especially in the case of having multiple gap-opening planets in the disc.

\section{Conclusions} \label{sec:conc}
This work demonstrates a scenario of sequential giant planet formation that is triggered by an initial disc substructure. We further extended the model in \cite{Lau2022} by including the effects of planetary gas accretion and gap opening. We employed DustPy to model a protoplanetary disc initially with micron-sized dust, and SyMBAp was employed to model the evolution of the planetesimals upon formation.

Consistent with the previous results, planetary cores are formed rapidly from the initial disc substructure, which can then be retained at the migration trap and start gas accretion. The results show multiple (up to three) cores can form and grow into giant planets in each generation. As the first generation of gas giants has formed and opened a steady gap, the new pressure bump at the outer edge of the planetary gap becomes the next location of planet formation.

In the case of the higher value of $\alpha=5\times10^{-4}$, the second generation of planet formation occurs about 1 Myr after the first one, and only ice giants were formed instead of gas giants. This case also shows a high planet formation efficiency where more than 85\% of the dust beyond 5 au is converted into massive bodies. As the first generation of gas giants effectively prevent dust from flowing through to reach the inner disc, the retained dust is then available for the next generation of planet formation. In the case of a lower value of $\alpha=3\times10^{-4}$, a larger degree of stochasticity was shown, while the general scenario of sequential giant planet formation remains. In both cases, a compact chain of giant planets are formed at the end of the simulations. While the simulations were stopped at 2 Myr, a natural continuation to the model would be to include the effect of photoevaporation to physically dissipate the disc and stop gas accretion.

Although the formation mechanisms of disc substructure are beyond the scope of this work, further investigations are required to study the possible shape and location produced by physical processes. It is unlikely that any disc substructure can trivially provide the conditions required for planetesimal or planet formation. Also, the parameter space and the number of random simulations in this work are limited by the computational costs. Further code optimisation is required to study the statistical effects and to model the diversity of planetary systems. And, planetary gas accretion is still an active field of research. Further investigations specifically on gas accretion are required to model the formation time of the Solar System's giant planets.

\begin{acknowledgements}
We thank T. Kleine, C. Ormel, R. Teague and A. Vazan for the insightful discussions.
We acknowledge funding from the European Research Council (ERC) under the European Union's Horizon 2020 research and innovation programme under grant agreement No 714769, under 
 the European Union's Horizon Europe Research and Innovation Programme 101124282 (EARLYBIRD) and funding by the Deutsche Forschungsgemeinschaft (DFG, German Research Foundation) under grant 325594231 and Germany’s Excellence Strategy – EXC-2094 – 390783311. JD was funded by the European Union under the European Union’s Horizon Europe Research \& Innovation Programme 101040037 (PLANETOIDS). Views and opinions expressed are however those of the authors only and do not necessarily reflect those of the European Union or the European Research Council. Neither the European Union nor the granting authority can be held responsible for them.
\end{acknowledgements}


\end{document}